\documentclass{iopart} 
\usepackage{iopams}  
\usepackage{graphicx,graphics,rotating}	
\graphicspath{{images/},{figures/}}
\usepackage[utf8]{inputenc}
\usepackage{eqnarray}
\expandafter\let\csname equation*\endcsname\relax 
\expandafter\let\csname endequation*\endcsname\relax
\usepackage{amsmath,amssymb,bbm} 

\def\notext#1{}
\newcommand{\KI}{\text{KI}}

\newcommand{\vd}{y}	
\newcommand{\hd}{x}

\renewcommand{\>}{\rangle}
\def\1{{\mathchoice{\rm 1\mskip-4mu l}{\rm 1\mskip-4mu l}{\rm 1\mskip-4.5mu l}{\rm 1\mskip-5mu l} }}

\newcommand{\ZZ}{\mathbb{Z}}

\newcommand{\ii}{ {\rm i} }

\newcommand{\bra}[1]{{\langle #1 \vert}}

\newcommand{\ket}[1]{{\vert #1 \rangle}}

\newcommand{\ave}[1]{{\langle #1\rangle}}

\newcommand{\appropto}{\mathrel{\vcenter{
  \offinterlineskip\halign{\hfil$##$\cr
    \propto\cr\noalign{\kern2pt}\sim\cr\noalign{\kern-2pt}} } }}

\usepackage{hyperref}

\providecommand{\openone}{\leavevmode\hbox{\small1\kern-3.8pt\normalsize1}}

\begin{document} 
\title{Two dimensional kicked quantum Ising model: dynamical phase transitions} 

\author{C Pineda$^{1}$, T Prosen$^2$, E Villaseñor$^{3}$}

\address{$^1$Instituto de Física, Universidad Nacional Autónoma de México, México D.F. 01000, México}
\address{$^2$Department of Physics, Faculty of Mathematics and Physics, University of Ljubljana, 1000 Ljubljana, Slovenia}
\address{$^3$Facultad de Ciencias, Universidad Nacional Autónoma de México, México D.F. 01000, México}

\date{\today}%
\begin{abstract}
Using an efficient one and two qubit gate simulator, operating on graphical processing units, we investigate ergodic properties of a quantum Ising spin $1/2$ model on a two dimensional lattice, which is periodically driven by a $\delta$-pulsed transverse magnetic field.  We consider three different dynamical properties: (i) level density and (ii) level spacing distribution of the Floquet quasienergy spectrum, as well as (iii) time-averaged autocorrelation function of components of the magnetization.  Varying the parameters of the model, we found transitions between ordered (non ergodic) and quantum chaotic (ergodic) phases, but the transitions between flat and non-flat spectral density {\em do not} correspond to transitions between ergodic and non-ergodic local observables.  Even more surprisingly, we found nice agreement of level spacing distribution with the Wigner surmise of random matrix theory for almost all values of parameters except where the model is essentially noninteracting, even in the regions where local observables are not ergodic or where spectral density is non-flat. These findings put in question the versatility of the interpretation of level spacing distribution in many-body systems and stress the importance of the concept of locality.
\end{abstract}
%
%
\maketitle
\section{Introduction}  

Quantum dynamics of strongly interacting quantum systems is one of the most
exciting fields of current physics research, in particular due to the fact that
many fundamental phenomena, such as thermalization and equilibration in large
closed systems~\cite{Polkovnikov2011} are still lacking fundamental
understanding. Moreover, as such models are very difficult to simulate by best
contemporary computers due to exponential growth of Hilbert space 
dimension~\cite{DeRaedt2007121},
theoreticians have very little predictive power as the theory of
non-equilibrium quantum thermodynamics is only beginning to
emerge~\cite{Gemmer,RevModPhys.82.277}.

On the more mathematical physics side, the key question is a precise
understanding of the notion of quantum ergodicity, and the ergodic to
non-ergodic transitions in thermodynamic limit. 
By ergodicity we mean that for most observables, the time average of an
observable and its time correlation will coincide with the canonical average of
the observable alone. For periodically driven quantum systems this means that
starting from almost any initial state, all observables and correlations are
determined by an infinite temperature Gibbs state.  Lack of ergodicity thereof
implies a sort of localization within a part of many body Hilbert space.
This type of many-body localization is essentially different and
perhaps more mysterious from the one studied in many-body systems with on-site
disorder \cite{altshuler,huse,znidaric,pollmann}.

Some conceptual and computational attempts in this direction have been
made in Refs.~\cite{Prosen1998, Prosen1998b, Prosen1999,Prosen2002, Polkovnikov2013a,
Polkovnikov2013b,rigol}.  In particular, one of us
showed~\cite{Prosen1998,Prosen1998b,Prosen1999,Prosen2002} that 
kicked $XX-Z$ chains and kicked Ising spin chains
typically exhibit non-ergodic to ergodic transition with increasing
period of the kicking, 
despite the fact that the limiting ({\em autonomous}) system when the
driving period goes to zero may be {\em integrable}.
This transition was accompanied with the transition in
(quasi)energy level spacing distribution going from Poissonian statistics of
uncorrelated levels for integrable regimes to Wigner-like level spacing
distribution accurately describing Gaussian orthogonal (or unitary) ensembles
of random matrix theory (GOE) for non-integrable regimes. 
The fact that in the semi-classical limit of small effective Planck constant (if the latter can be meaningfully defined), both transitions (in the spectral
correlations and the ergodicity of the system) are in one to one
correspondence~\cite{haakebook}, has lead to an intuitive belief that that
should hold in more general cases, such as many body quantum systems without a small $\hbar$ parameter. This,
however, has not been carefully explored yet.
Yet a third measure of
quantum chaos or quantum ergodicity in periodically driven quantum systems can
be introduced, going under different names as {\em Loschmidt echo}, {\em
survival amplitude}, or equivalently, (Fourier transform of) the quasienergy
spectral density. We shall simply refer to it as the {\em spectral density},
and for ergodic kicked quantum systems one may expect it to be constant.
Recently it has been suggested that for quantized chaotic single particle
periodically kicked systems (e.g., for the so-called kicked top), one obtains
dynamical instabilities signalled by discontinuous (sharp) transitions, or
singularities in the spectral density. We shall argue later that spectral
density is typically always constant in kicked locally interacting quantum spin
$1/2$ chains, so there could be no transitions, but an interesting question
opens of what happens in higher dimensional spin lattices.

In this paper we study probably the simplest non-trivial dynamical many-body
model in two dimensions, namely the two-dimensional version of kicked Ising
spin $1/2$ model introduced in Ref.~\cite{Prosen2002}. We numerically
accurately calculate different dynamical and spectral properties of the model on
a finite periodic rectangular $L_x \times L_y$ lattices implementing local one
and two qubit operations on graphical processing units (GPUs) of a desktop
computer, in a spirit similar to~\cite{2013arXiv1305.0036D}. Namely we compute
phase diagrams (dependencies on the model's parameters) of dynamical
susceptibilities, spectral densities and quasienergy
level spacing distributions. We find that in most parts of parameter space
these quantities turn out to be stable against increasing the lattice sizes, so
we formulate certain conjectures about the thermodynamic behavior.
We find, firstly, similarly as in one-dimensional
chains, well defined transitions between non-ergodic and ergodic dynamical
susceptibilities of local observables.
Secondly, we find non-trivial transitions between flat spectral density and
spectral densities with the shapes essentially determined by a single Fourier
mode, $\rho(\phi) = \frac{1}{2\pi} + c \cos( k\phi)$, for some constant $c$ and integer $k$. 
The Fourier coefficient $c$ seems to be decreasing with increasing the lattice size, but for any fixed
Hilbert space dimension one finds a dramatic difference of $|\rho(\phi)-\frac{1}{2\pi}|$ from a prediction
of a flat spectrum (say of a random unitary matrix).
However, remarkably and surprisingly the points of transitions of dynamical
susceptibilities {\em do not} correspond to the points of transitions of level
density.  Thirdly, analysis of level spacing distributions of properly
unfolded \cite{haakebook} quasienergy spectra reveals universal GOE (Wigner)
statistics across most of the parameter space whereas non-universal statistics are
only observed on singular parameter regions with trivially integrable dynamics.
This suggests that there are no nontrivial integrable points.
Therefore the transition in spectral correlations is instantaneous in
thermodynamic limit and does not correspond with any other measure of quantum
ergodicity, such as dynamical susceptibilities and spectral densities. We
believe this is a remarkable observation which can put under question the
versatility of level spacing distribution in quantum many-body systems. In
particular, we believe other (direct) measures of quantum ergodicity should be
used when discussing thermalization or equilibration.

Our paper is organized as follows. In section 
\ref{sec:model} we introduce the model, and discuss its general properties, including its symmetries. 
In section~\ref{sec:density} we comment on the spectral density, and observe its properties
for all values of the parameters of the model. A first picture of the model emerges. 
In the next part (section~\ref{sec:ps}), we discuss the correlation properties of the spectra, using 
the nearest neighbour spacing distribution. The second picture emerges. We next (section~\ref{sec:ct})
study the dynamical susceptibilities (autocorrelation functions of certain observables) where a third
picture of the model arises. Finally, we gather the results to come up with our conclusions
in section~\ref{sec:conclusions}.
\section{The two dimensional quantum kicked Ising model} 
\label{sec:model}


We study a periodic 2 dimensional lattice of kicked spin $1/2$ particles, 
inspired by the Kicked Ising (KI)
chain, proposed by one of the authors in \cite{Prosen2002}.
The particle at site $m \in \{1,\ldots,L_x\}$,
$n \in \{1,\ldots,L_y\}$ will be described by standard Pauli
operators $\sigma^\alpha_{m,n}$, $\alpha\in\{x,y,z\}$. 

Let us start by defining a 2 dimensional Ising Hamiltonian 
\begin{equation}
H_1 = J H_{\rm I},\quad 
   H_{\rm I}=\sum_{m=0}^{L_x-1}\sum_{n=0}^{L_y-1} 
     (\sigma^z_{m,n}\sigma^z_{m+1,n} + \sigma^z_{m,n}\sigma^z_{m,n+1}),
\end{equation}
with periodic boundary conditions
$\sigma^\alpha_{m,L_y}\equiv
\sigma^\alpha_{m,0}$, $\sigma^\alpha_{L_x,n}\equiv \sigma^\alpha_{0,n}$.
We now define a Zeeman Hamiltonian for a spatially homogeneous magnetic field
$\vec{b}$ 
\begin{equation}
H_0 = \sum_{m=0}^{L_x-1}\sum_{n=0}^{L_y-1} \vec{b}\cdot\vec{\sigma}_{m,n} 
       = \vec{b}\cdot \vec{S},\quad \vec{S} 
       =: \sum_{m=0}^{L_x-1}\sum_{n=0}^{L_y-1} \vec{\sigma}_{m,n}.
\end{equation}
Notice that we can always choose the coordinate system such that
$\vec{b}=(b_x,0,b_z)$, so that both $H_0$ and $H_1$ are real. We will normally
consider only a transverse field, that is, $\vec{b}=(b_x,0,0)$. 
The parameters, $J$ (inter-spin interaction) and $b_x$ (transverse magnetic
field) are independent dimensionless parameters that specify the model.

We consider a time-dependent Hamiltonian, where the magnetic field is modulated
by periodic $\delta-$pulses of period $\tau$
\begin{equation}
H(t) = H_1 + H_0 \sum_{j\in\ZZ} \delta(t-j\tau).
\end{equation}
One-step quantum evolution propagator for the KI model over one period of the model
$U(t)={\cal T}\exp\left(-\ii \int_{0^+}^{1^+} {\rm d}t H(t)\right)$ ---
the so-called Floquet map --- reads, setting $\tau=1$ by a free choice
of units:
\begin{equation}
U_{\rm KI} = U_{\rm Ising}(J) U_{\rm kick}(\vec{b}),
\label{eq:floquet}
\end{equation}
where
\begin{equation}
U_{\rm Ising}(J)  = \exp(-{\rm i}H_1) =\exp(-\ii J H_{\rm I}),\quad
U_{\rm kick}(\vec{b}) = \exp(-{\rm i}H_0)=  \exp(-\ii \vec{b}\cdot \vec{S}).
\end{equation}
\subsection{Symmetries} 
\label{sec:symmetries}
We shall now briefly discuss some obvious symmetries of the model which help in
reducing computational complexity of simulations.

\paragraph{Parameter space symmetries.} 
Notice that the system is periodic in the parameters 
since the spectra of operators $H_{\rm I}$ and $\hat{b}\cdot\vec{S}$, where
$\hat b = \vec{b}/|\vec{b}|$ is the unit vector in the direction of $\vec b$,
form subsets of integers.
In fact, in the spectrum of $H_{\rm I}$ there can be only integers with fixed remainder of division with $4$, since flipping an arbitrary spin can change $H_{\rm I}$ only by $\pm 4$ or $\pm 8$. 
More precisely, the spectrum of $H_{\rm I}$ consists of points
$\{ 2 L_x L_y, 2 L_x L_y - 4, 2 L_x L_y - 8,\ldots, -2 L_x L_y\}$, hence
\begin{equation}
U_\text{Ising}(J+\pi/2) = (-1)^{L_x L_y} U_\text{Ising}(J).
\end{equation} 
Similarly, the spectrum of $\hat{b}\cdot\vec{S}$ consists of points $\{ L_x L_y, L_x L_y - 2, L_x L_y - 4,\ldots,-L_x L_y\}$, hence
\begin{equation}
U_\text{kick}(\vec{b}+\pi \hat b) = (-1)^{L_x L_y} U_\text{kick}(\vec{b}).
\end{equation}
Let us now further assume that the field is transverse $\hat{b}=(1,0,0)$ as will be the case for most of this paper.
Then, performing a checkerboard canonical (unitary) transformation $\hat{C}$, namely: flipping the signs of $y,z$ components of spins $\vec{\sigma}_{m,n}$ for
all even $m+n$ (i.e. rotating for angle $\pi$ along the $x$-axis), one finds that 
\begin{equation}
U_\text{Ising}(-J)U_\text{kick}(\vec{b}) = \hat{C}^{\dagger} U_\text{Ising}(J)U_\text{kick}(\vec{b}) \hat{C}.
\end{equation}
Similarly, canonical transformation $\hat{D}$, which flips $x,y$ components of {\em all} spins (rotates around $z$ axis for angle $\pi$) yields
\begin{equation}
U_\text{Ising}(J)U_\text{kick}(-\vec{b}) = \hat{D}^{\dagger} U_\text{Ising}(J)U_\text{kick}(\vec{b}) \hat{D}.
\end{equation}
Therefore, changing the sign of $J$ or $b_x$ leaves invariant all physical properties of the model, in particular the spectrum of $U_{\rm KI}$, so the principal domain of the phase diagram of the transverse field KI model only consists of a rectangle $(J,b_x) = [0,\pi/4]\times [0,\pi/2]$.
\paragraph{Symmetry reduction of the Hilbert space.} 
In the general case, the system is symmetric under the following geometric operations, generating the symmetry group of the model:
reflexion over the horizontal axis $R_\hd$, 
reflexion over the vertical axis $R_\vd$, 
horizontal translation $T_\hd$, and 
vertical translation $T_\vd$.

To illustrate these symmetries, let us number 
the sites in 
a $(L_\hd=4) \times (L_\vd = 3)$ grid from left to right, and bottom to top, 
and consider a state of the computational basis 
$|\psi_0\> = | i_0 i_1 \cdots i_{11} \>$, with $i_j \in \{0,1\}$. 
The action of $R_\hd$ on the grid will be to transform it into 
\begin{equation}
\begin{matrix}
8 & 9 & 10 & 11 \\
4 & 5 & 6 & 7 \\
0 & 1 & 2 & 3
\end{matrix}
\qquad \to  \qquad
\begin{matrix}
0 & 1 & 2 & 3 \\
4 & 5 & 6 & 7 \\
8 & 9 & 10 & 11 
\end{matrix},
\end{equation}
so
\begin{eqnarray*}
R_\hd|\psi_0\> & = | i_8 \cdots i_{11} i_4 \cdots i_7 i_0 \cdots i_{3} \>. 
\end{eqnarray*}
The action of the reflection $R_\vd$ is similar:
\begin{equation}
\begin{matrix}
8 & 9 & 10 & 11 \\
4 & 5 & 6 & 7 \\
0 & 1 & 2 & 3
\end{matrix}
\qquad \to  \qquad
\begin{matrix}
 11 & 10& 9 & 8  \\
 7  & 6 & 5 & 4 \\
 3  & 2 & 1 & 0 
\end{matrix},
\end{equation}
and the action over a member of the computational basis is 
\begin{eqnarray*}
R_\vd|\psi_0\> & = | i_3 \cdots i_{0} i_7 \cdots i_4 i_{11} \cdots i_{8} \>.
\end{eqnarray*}
Similarly we can picture the effects of the translations. The effect 
of vertical and horizontal translations on the original grid are
\begin{equation}
\begin{matrix}
8 & 9 & 10 & 11 \\
4 & 5 & 6 & 7 \\
0 & 1 & 2 & 3
\end{matrix}
\to  
\begin{matrix}
4 & 5 & 6 & 7 \\
0 & 1 & 2 & 3 \\
8 & 9 & 10 & 11
%
\end{matrix}, \qquad
\begin{matrix}
8 & 9 & 10 & 11 \\
4 & 5 & 6 & 7 \\
0 & 1 & 2 & 3
\end{matrix}
\to  
\begin{matrix}
11 & 8 & 9 & 10 \\
7  & 4 & 5 & 6 \\
3  & 0 & 1 & 2 
\end{matrix},
\end{equation}
respectively. Thus, the action of the symmetry  is simply 
\begin{eqnarray*}
T_\hd|\psi_0\> & = | i_8 \cdots i_{11} i_0 \cdots i_7 \>,
\text{ and}\\
T_\vd|\psi_0\> & = | i_3 i_0 i_1 i_2 i_7 i_4 i_5 i_6 i_{11} i_8 i_9 i_{10}\>.
\end{eqnarray*}
Notice that $R_\vd^2 =R_\hd^2 = T_\vd^{L_\vd} =   T_\hd^{L_\hd} = \openone$.
It can also be noted that $T_\vd = T^{L_\hd}$, where $T$ is the translation 
operator that acts as $T_|\psi_0\> = | i_{11}i_1 i_2 \cdots i_{10} \>$. 
The symmetry subspaces of the Hilbert space are therefore specified by two quasi-momenta $k_x \in \{0,\ldots,L_x-1\},k_y\in \{ 0,\ldots,L_y-1\}$, and for symmetric sectors with $k_{x,y}=0$ or $k_{x,y}=L_{x,y}/2$ by additional reflection signs $\pi_x,\pi_y\in\{\pm 1\}$, such that the states $|\psi\>$ from the subspace satisfy
$R_{x,y} | \psi\> = \pi_{x,y} |\psi\>, T_{x,y} |\psi\> = e^{-2\pi\ii k_{x,y}/L_{x,y}}|\psi\>$.

When we consider the special case of a transverse magnetic field, another symmetry arises. 
A parity operator $\prod_{m,n} \sigma^x_{m,n}$ commutes with the Floquet operator:
$[\prod_{m,n}\sigma^x_{m,n}, U_\KI]=0$. 

Our KI model also has an {\em anti-unitary symmetry} namely if $K$ is a complex conjugation in the standard Pauli basis, then
$K^{-1}\sigma^z_{j,k} K = \sigma^z_{j,k}$, 
$K^{-1}\sigma^x_{j,k} K = \sigma^x_{j,k}$, and
$K^{-1}\sigma^y_{j,k} K = -\sigma^y_{j,k}$. Writing a symmetrized Floquet propagator $U_{\rm KI}' = e^{-\ii H_0/2} U_{\rm KI} e^{\ii H_0/2} = e^{-\ii H_0/2} e^{-\ii H_1} e^{-\ii H_0/2}$ we then have immediately
\begin{equation}
K^{-1}U'_{\rm KI} K = (U'_{\rm KI})^{-1}, \quad {\rm and}\quad U'_{\rm KI} = (U'_{\rm KI})^T.
\end{equation}
Using the standard wisdom \cite{haakebook}, the model should then -- if `quantum chaotic' -- correspond to Circular orthogonal ensemble (COE) of random unitary symmetric matrices.

\subsection{Steady field limit} 
With the same machinery we can study the time independent limit,
corresponding to keeping $\vec b/J$ fixed, while letting $|J|$ go to zero.
Then, the scaled Hamiltonian is simply 
\begin{equation} \label{eq:time:independent:hamiltonian}
	H=H_1+ H_0.
\end{equation}
In order to study this Hamiltonian we also used the CUDA machinery, and used both first 
and second order Trotter approximations. For the results presented here, we
verified that the first order Trotter evolution gives essentially the same results as 
the second order, meaning that the changes in the figures presented are so small 
that cannot be noticed. 
\section{Spectral density} 
\label{sec:density}

The spectrum of the Floquet map, ${\cal S}=\{ \phi_n; n=1,\ldots,{\cal
N}:=2^{L_x L_y} \}$, defined by the unitary eigenvalue problem
\begin{equation}
U_{\rm KI}\ket{\psi_n} = e^{-\ii \phi_n}\ket{\psi_n},
\end{equation} 
entails the main
dynamical features of the model. The statistical properties of ${\cal S}$ for
systems with chaotic classical limit has been the central theme
of quantum chaos~\cite{haakebook}. However, very little is
known about the distribution of $\{\phi_n\}$ for many-body quantum models,
despite the fact that full many-body quantum dynamics is becoming
experimentally accessible in recent years, in particular in cold atom
laboratories~\cite{Britton2012}.  Even the behaviour of the simplest  spectral
characteristic, the 1-point function or the spectral density, defined as
\begin{equation}
\rho(\phi) = \frac{1}{\cal N} \sum_{n=1}^{\cal N} \delta(\phi-\phi_n),
\label{rho}
\end{equation}
would be of great interest to know.  
In autonomous (time-independent) quantum
many-body systems the spectral density is predicted to go to a Gaussian in
thermodynamic limit~\cite{mahler,bogomolny}, while for periodically
driven quantum systems one may perhaps intuitively expect (and observe, in 1D
chains~\cite{Prosen1999}) that the Floquet quasienergy spectral density would
be the constant (flat) function $\rho(\phi) = \frac{1}{2\pi}$, in a generic
case.

\begin{figure} 
	\begin{center} 
  \includegraphics[scale=0.85]{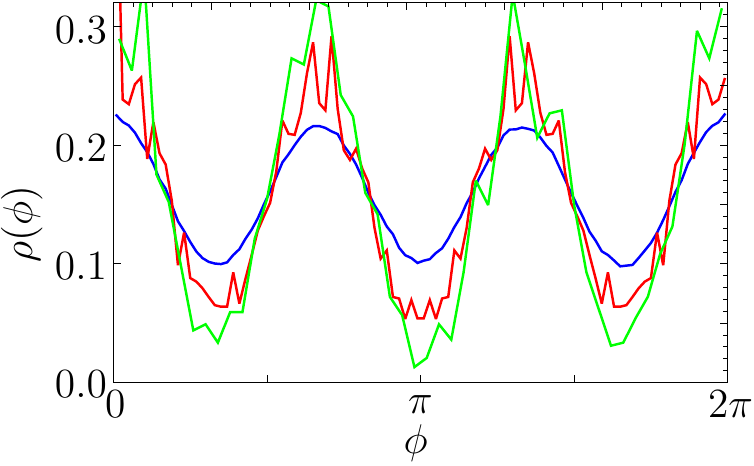}\hfill%
  \includegraphics[scale=0.85]{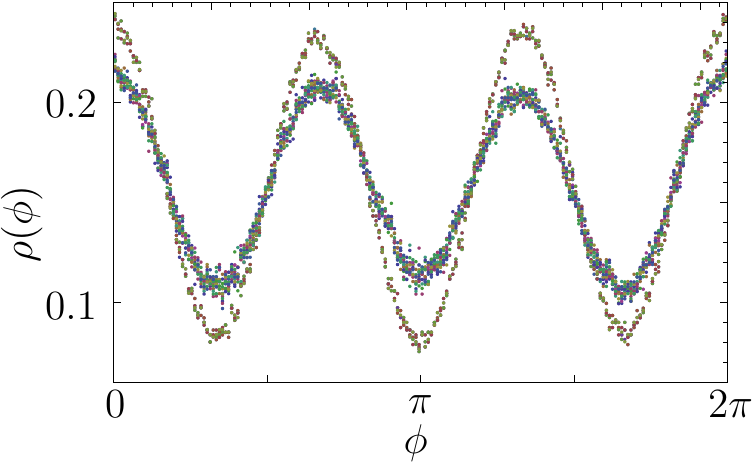}
	\end{center}
	\caption{(left panel) Spectral density as a function of the number of
	particles, for $b_x=0.2$ and $J=0.5$. The different sizes are $3\times 4$ (green),
	$4\times 4$ (red) and $5\times 4$ (blue).  
	(right panel) Spectral density for a $5\times 4$ lattice and the same values
	of $b_x$ and $J$ as the figure on the left. Each symmetry  sector is
	coded in a different (but not specified) color. The data oscillating with a bigger amplitude correspond to 'generic' symmetry sectors, while the data oscillating with a smaller amplitude correspond to symmetric         sectors with $k_{x,y} =0$ or $k_{x,y} = L_{x,y}/2$. }
	\label{fig:spectral:density}
\end{figure} 

The spectral density $\rho(\phi)$ is a $2\pi$-periodic function and therefore can be
represented in terms of the Fourier modes as
\begin{equation}
\rho(\phi) = \frac{1}{2\pi}\left(1 + 2 \sum_{k=1}^\infty \rho_k \cos(k\phi)\right),
\label{cosine}
\end{equation}
where the Fourier coefficients $\rho_k$ are given as traces of the $k-$step KI propagator
\begin{equation}
\rho_k := \int_{0}^{2\pi}\!\! {\rm d}\phi\, \rho(\phi) e^{\ii k \phi} = \frac{1}{\cal N}\tr U_{\rm KI}^k.
\label{rhok}
\end{equation}
The symmetry property $\rho_{-k} = \rho_k$ has been used, following from the
symmetry of the spectra of $H_0$ and $H_1$ around zero energy and cyclicity of
the trace.

\begin{figure} 
	\begin{center} 
  \includegraphics[scale=1]{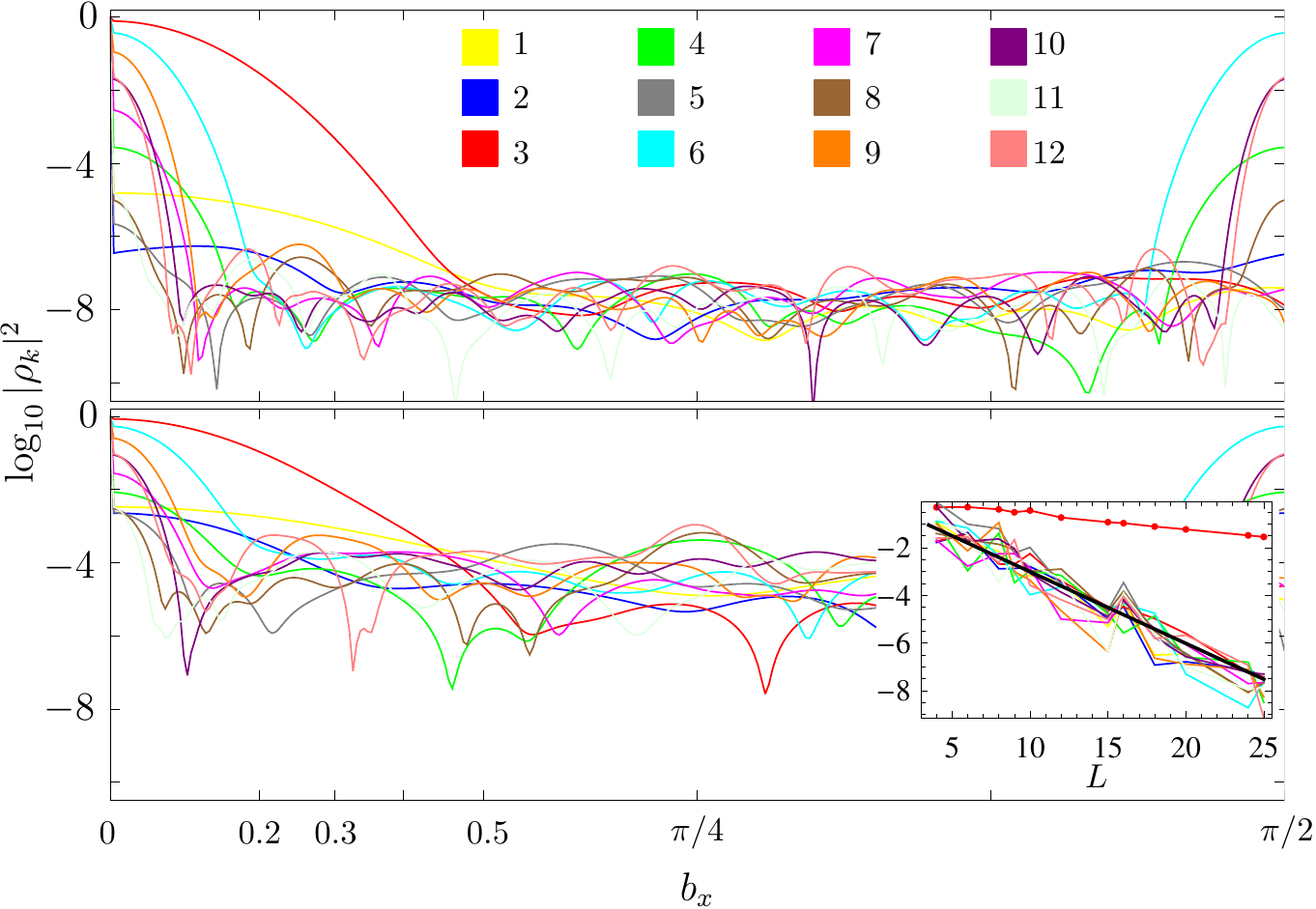}
	\end{center}
	\caption{We plot different Fourier components $\rho_t$, $t=1,\cdots,12$
	(color coded) in a semilogarithmic plot, varying the magnetic field
	$b_x$ (in the main plots) and varying the size (in the inset) with a
	fixed $J=0.5$.  On the top panel a $5\times5$ lattice is studied,
	whereas in the lower panel a $4\times 4$ lattice is used.
	In the inset $\log_{10} |\rho_3|$ for $b_x=0.2$ is shown (as a red line
	with points), together with all $|\rho_{1,\ldots,12}|$ components
	for $b_x=0.6$ (as simple broken lines). A thick black line corresponding
	to $|\rho| \propto N^{-1/2}$ is also shown for
	comparison.} \label{fig:transition:for:two:phases}
\end{figure} 
The sum in (\ref{rho}) can be carried in the whole Hilbert space, or in a
single symmetry sector (with fixed quasi momenta and/or parities).  We
calculated the spectra directly, using a basis that splits the evolution
operator into different sectors, and numerically fully diagonalizing each sector
independently. That way, we could calculate 
the whole spectrum for sizes of up to $5\times4$.  The behavior
over different symmetry sectors seems to be similar, in all examples that we
considered, see Fig.~\ref{fig:spectral:density} (right panel).

For efficient numerical computation of the leading Fourier components $\rho_k$,
one should instead use the expression in terms of traces of powers of the propagator
directly (\ref{rhok}). The computation can be further simplified by noting that the
trace over the many-body Hilbert space is self-averaging and can be approximated
using an expectation value in a single typical (random) state. Namely
\begin{equation}
\frac{1}{\cal N}\tr U_{\rm KI}^t = \int \bra{\psi}U_{\rm KI}^t\ket{\psi} \rmd \mu(\psi) 
\approx  \bra{\psi_\text{random}}U_{\rm KI}^t\ket{\psi_\text{random}},
\label{eq:approximation:traceUt}
\end{equation}
where ${\rm d}\mu(\psi)$ is the measure induced by the Haar measure over
the unitary group, and $\ket{\psi_\text{random}}$ is a state drawn at
random with the Haar measure. Moreover, if one selects that state belonging to
a given symmetry subspace, one then studies the spectral properties of that particular
subspace.  For practical computation, one may take all components of
$\ket{\psi}$ as random Gaussian $c-$numbers with zero mean and equal variance
and then normalize the state.  In our case, calculating the kick-by-kick
evolution of a state is quite efficient, so this form of calculating the
Fourier transform of the spectral density is particularly convenient. 


We now examine the behaviour of the spectral density, for several parameter
values, and several sizes. The left panel of fig.~\ref{fig:spectral:density}
suggests a dominant Fourier component $\rho_k$ of the spectral density, 
whose magnitude varies with the size of the system. Examining each of the 
Fourier contributions as a function of the parameters proofs very useful. 
Such analysis is carried out for all coefficients up to $k=12$, varying 
the transverse component of the magnetic field. A similar behaviour 
is obtained for different sizes as can be appreciated in
\fref{fig:transition:for:two:phases}. 
%
%
%
The most outstanding fact is that there are clearly two different regions.  One
in which we have all Fourier coefficients magnitude close to the average random
value, given by $|\rho_\text{noise}|$, and another region, in which there is an
ordered phase, manifested by $|\rho_k| \gg |\rho_\text{noise}|$.  We estimate
$|\rho_\text{noise}|$, replacing $U_{\rm KI}^t$ in
\eref{eq:approximation:traceUt} by a random unitary matrix of dimension $2^{L_x L_y}$. 
This gives rise to $|\rho_\text{noise}| \propto 2^{-L/2}$ where $L=L_x L_y$,
which is in good agreement with the tendency
observed.  We have also plotted the dominant Fourier component at $b_x=0.2$.
Indeed, there is a small decay of the oscillations for the $b_x=0.2$, however,
the comparative effect
enhances with the system size, in the sense that $\left| \rho_3(b=0.2) /
\rho_\text{noise}\right| \appropto\exp(0.27 L)$ thus sharpening the transition
from a disordered to an ordered phase. Therefore, even though the spectral
density $\rho(\phi)$ seems to universally approach a constant $1/(2\pi)$ when
one approaches the thermodynamic limit $L_x,L_y\to\infty$, there are
discontinuous transitions on the size scaling of the deviation
$\rho(\phi)-\frac{1}{2\pi}$ with changing the system parameters.

%
%


\begin{figure} 
	\begin{center} 
 	\includegraphics[width=\columnwidth]{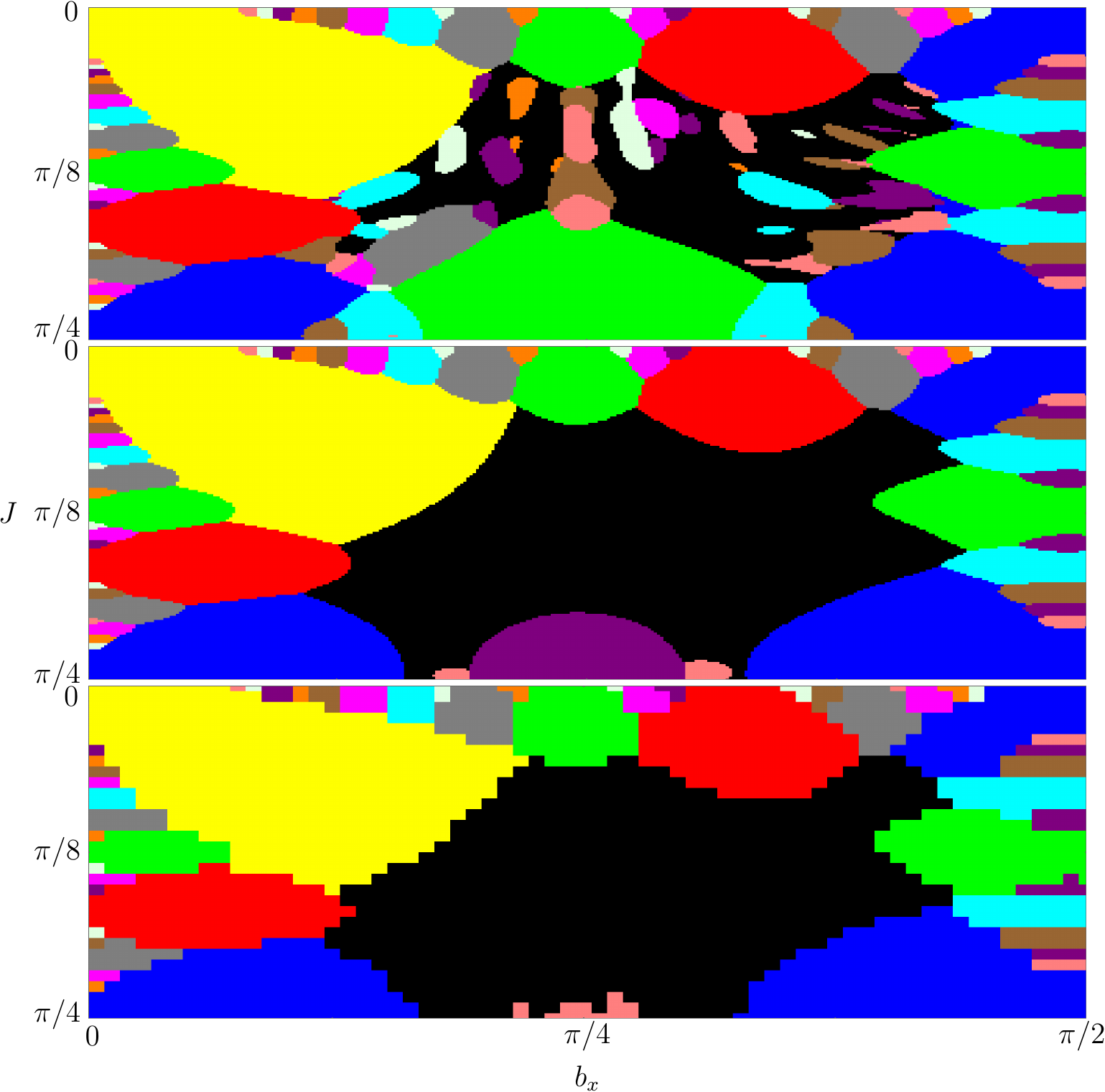} 
	\end{center}
	\caption{Phase diagram of the spectral density.  We color code the
	different dominant Fourier components, according to the coding
	displayed in the legend of \fref{fig:transition:for:two:phases} (see
	main text for details).  We show diagrams for different dimensions;
	$4\times 3$, $4\times 4$ and $ 4\times 5$, from top to bottom,
	respectively. }
	\label{fig:all:phases}
\end{figure} 

One can get an interesting global picture of the model by plotting a spectral
density phase diagram.  Namely, we determine and plot the leading nontrivial
spectral component $k$ for which $\rho_k$ is dominating, as a function of
model's parameters.  We shall consider the spectrum to be {\em flat}, $k=0$, if
all Fourier coefficients $\rho_k$, for $k > 0$, are comparable to 
$\rho_\text{noise}$.  
That is, the spectrum is declared flat, if
\begin{equation}
\rho_k <  20 \rho_\text{noise}, \quad \forall k > 0 .
\label{eq:criterion:flat}
\end{equation}
On the other hand, we consider the system to be in the phase $k > 0$, if
$|\rho_k| > |\rho_{k'}|$,  for all $k \ne k' \neq 0$.  We note that typically a
single Fourier component is dominating others by several orders of magnitude.
The gap between the Fourier components even increases when we increase the
lattice size (see \fref{fig:transition:for:two:phases}).  See
Figure~\ref{fig:all:phases} for a comparison of phase diagrams for different
lattice sizes which seems remarkable stable.

\section{Level spacing distribution} 
\label{sec:ps}
\begin{figure} 
	\begin{center} 
	\includegraphics[width=\columnwidth]{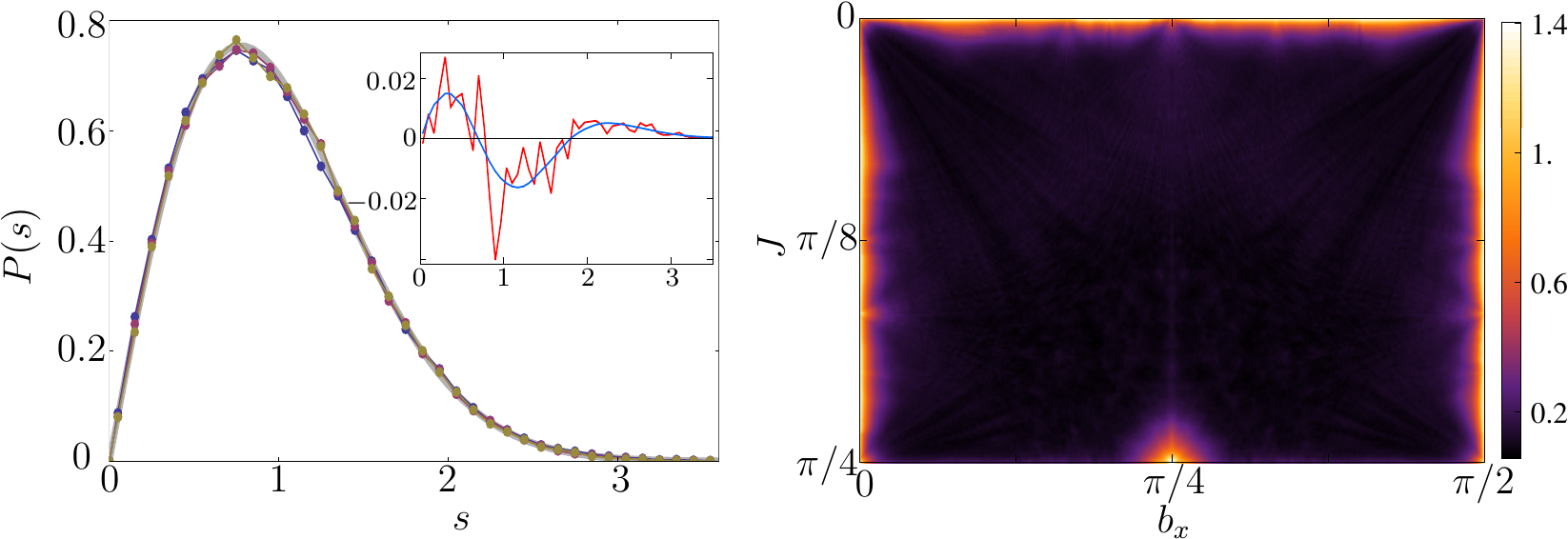} 
	\end{center}
	\caption{Analysis of the distribution of the nearest neighbour spacing
	$P(s)$. On the left panel, we observe the nearest neighbour spacing distribution for three
	different transverse fields, $b_x=0.2$, $0.3$ and $0.5$ in  red, green
	and yellow respectively,  $J=0.5$, and we consider a $5\times 4$
	lattice.  In all cases, we are considering $s_x = \pm 1$, $k_x \in
	\{1,2\}$ and $k_y=1$.  The thick black curve correspond to the Wigner
	surmise.  In the inset, we show the average of these three curves,
	minus the Wigner surmise, together with the theoretical prediction. 
	On the right panel, we consider the Kolmogorov distance between 
	the unfolded $P(s)$, and the Wigner surmise, for all the parameters
	of the model, and a $4\times3$ lattice. Very good agreement with the
	RMT prediction is observed except when $J$ or $b_x$ are zero, or
	$J=b_x=\pi/4$. 
	}
	\label{fig:ps}
\end{figure} 
In order to analyze the correlation properties of the spectrum we used
the commonly studied nearest neighbour spacing
distribution. 
That is, we consider the distribution of level spacings 
\begin{equation}
\tilde s_i = \phi_{i+1} - \phi_i
\label{eq:def:unfolded:s}
\end{equation}
where $\phi_i$ are the sorted eigenphases of the evolution operator. 
In order to remove the effect of non-uniform level density we perform {\em
unfolding}, i.e., a smooth non-linear scaling of the eigenvalues in order to
get a uniform spectral density.  Since the density is typically well described
by few Fourier components, we shall take 6 of them to numerically perform the unfolding. 
Thus, we shall use the mapping 
\begin{equation}
\varphi_i = 2\pi \left( \frac{\phi_i}{2\pi}+ \sum_{k=1}^6 \alpha_k \frac{\sin k \phi_i}{k} \right)
\label{eq:unfolding}
\end{equation}
where  $\alpha_k = \rho_k/\pi $ is determined numerically using direct
evolution.  In this way we can unfold the spectra to obtain fairly flat
distributions of unfolded level spacings. Another very important aspect that
must be taken into account is the fact that different symmetry sectors are not
statistically correlated with each other, so we must rather look at the
distribution of 
\begin{equation}
s_i^K = \varphi_{i+1}^K -\varphi_i^K
\label{eq:good:definition:s}
\end{equation}
where $K=\{k_x, k_y, \pi_x, \pi_y, \Pi \}$ denotes the set of quantum numbers
that determines the irreducible quantum sector. 

By construction, the mean spacing $s_i^K$ equals one, so the probability
density $P(s)$ of $\{ s_i^K\}$ is normalized such that $\int_0^\infty P(s) {\rm
d}s = \int_0^\infty s P(s) {\rm d}s = 1$. 
The famous quantum chaos conjecture states that $P(s)$ behaves generically 
 as the corresponding classical ensemble of random matrices~\cite{haakebook},
 given that the classical limit is strongly chaotic (i.e., hyperbolic dynamical
 system). Given the time reversal symmetry, the corresponding ensemble, in our
 case, would be the circular orthogonal ensemble (COE).
%
To
a very good approximation, the level spacing distribution is given in terms of
$2\times 2$ random real symmetric matrices, the so-called Wigner surmise $P_{\rm Wigner}(s) =
\frac{\pi}{2} s \exp(-\pi s^2/4)$. A similar conjecture has been suggested for
strongly non-integrable quantum many-body systems \cite{guhr98random}, but it has not
been established precisely (yet), how non-integrability and level statistics 
are related in a given class of models.

We present in the left panel of \fref{fig:ps} the $P(s)$ for three typical
cases of our KI model, each of which behaves completely differently with
respect to the dynamical ergodicity measures discussed in this paper (spectral
density and dynamical susceptibility). We see, however, that $P(s)$ is in all
three cases excellently described by COE or Wigner surmise. In finer scale
(inset of the
left panel of \fref{fig:ps}) even the difference between Wigner's surmise and
the exact COE result can be resolved for the dynamical data.  In the right panel
of \fref{fig:ps} we plot the Kolmogorov distance, $\int_{\mathcal X} |f(x)
-g(x)| {\rm d} x $, between the observed distribution of nearest neighbour
spacings and the RMT prediction. There is good agreement with COE in the
{\em entire} parameter space except for trivial integrable cases of zero field
or zero spin interaction (both modulo $\pi/2$), or specially commensurate fields
where the spectrum of $U_{\rm KI}$ can be explicitly computed in terms of
regular or number-theoretic functions.

\section{Dynamical susceptibilities and non-ergodicity to ergodicity transition} 
\label{sec:ct}

So far we have analyzed dynamical properties of the 2D kicked Ising model
which depend solely on its spectrum.  Now we shall focus on {\it dynamical
correlations} of {\it local observables}, which are the key input to any linear
response treatment of condensed matter theory~\cite{kubogreen,
kubokubo, evans2007statistical}.

\begin{figure} 
	\begin{center} 
	\includegraphics[width=.6\columnwidth]{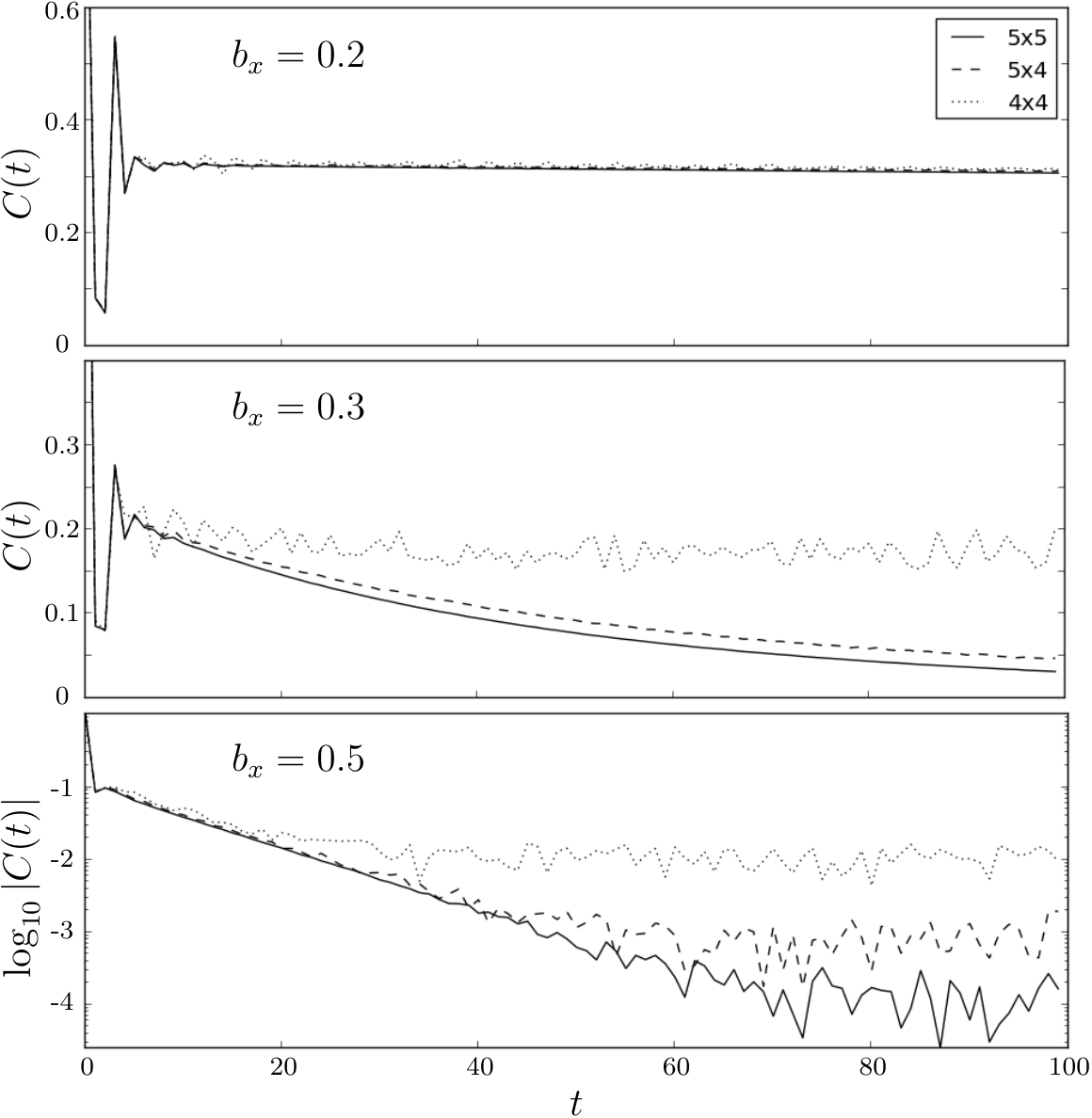} 
	\end{center}
	\caption{Correlation decay for the transverse field KI model, varying
	$b_x$, for different dimensions and fixed $J=0.5$. The calculation is done
	using a single random state.}
	\label{fig:correlacion:decay}
\end{figure} 

Consider a traceless observable $M$ (typically extensive and
local), say a
component of magnetization $M=S^\nu$, $\nu\in\{x,y,z\}$. We define its time-autocorrelation with
respect to 
the kicked Ising dynamics as
\begin{equation*}
C(t) = \frac{1}{L_x L_y {\cal N}} \tr( M(t) M )
\approx \frac{1}{L_x L_y} \bra{\psi_{\rm random}}M(t) M \ket{\psi_{\rm random}}.
\end{equation*}
If $\tr M \ne 0$, the
corresponding constant has to be subtracted from $C(t)$.
Here  $M(t)$ denotes the time-dependent observable in the Heisenberg picture: 
$M(t) = U_\KI^{-t} M U_\KI^t$, $t\in\ZZ$.  

One measures the ergodicity of an observable $M$ by the so-called dynamical
susceptibility, defined as the time-average of $C(t)$:
\begin{equation}
\chi_M = \lim_{T\to\infty} \frac{1}{T} \sum_{t=1}^T C (t).
\end{equation}
By definition, observable $M$ is {\em ergodic} with respect to dynamics $U_{\rm KI}(t)$, if
$\chi_M = 0$, and non-ergodic otherwise. 
Note that $\chi_M$ is always nonnegative as it represents the spectral weight,
i.e. the power spectrum of $M(t)$, at frequency $\omega=0$.

For numerical investigations, in order to diminish the transient effects of
relaxation, it is useful to define a finite time average between two,
sufficiently large times
$T_1,T_2\in\ZZ$,
as 
\begin{equation}
\ave{C(t)}_{T_1}^{T_2} = \frac{1}{T_2-T_1+1}\sum_{t=T_1}^{T_2} C(t).
\end{equation} 
In order to illustrate the ergodic properties of the model for different
parameters, we have analysed a series of observables. We used both, observables
symmetric under particle permutation, and non-symmetric observables, but restricted
ourselves to sums of few-site local observables.

%

We studied the general case with $M=\hat{n}\cdot\vec{S}$ with $\hat{n}$ an arbitrary
unit vector. Small system sizes revealed that when $\hat{b}$ points in 
any of the three axis directions ($x$, $y$, or $z$), the dynamical susceptibilities are {\em exactly} symmetric in parameter space with respect to the line $b_x=\pi/4$ in parameter
space.  However, this is the case for general observables, although there 
is a strong tendency to be exactly symmetric. More general observables also 
have this tendency, that becomes increasingly more difficult to explore
for moderate large systems.

We found three qualitatively different kinds of behaviour of dynamical
susceptibility, exemplified in \fref{fig:correlacion:decay}.  Here we compare
the behaviour for $M=S^x$ for different sizes of the system and a fixed
Ising interaction of $J=0.5$.  For $b_x=0.2$,
there is no decay neither for large times nor dimensions. There seems to be a
nonvanishing asymptotic value $\chi_M \neq 0$, which is also characteristic of
integrable systems. For $b_x=0.3$, $C(t)$ seems to decay algebraically to an
asymptotic value which decays with increasing lattice size, so it is
reasonable to conclude $\chi_M =0$.  
Finally, there are parameter values (say $b_x=0.5$) for which fluctuations
on top of an asymptotic
value, are reached exponentially fast. Again the asymptotic value decreases
with the Hilbert space dimension, as well as the fluctuations, suggesting
they both vanish in the thermodynamic limit. In fact, the sources of data
fluctuations at large times are twofold: finite size effects, and random
initial state sampling (approximating the trace), whereas empirical evidence
suggests that the latter (contributing to fluctuations as $\sim 1/\sqrt{2^{L_x
L_y}}$) quickly becomes negligible.
\begin{figure} 
	\begin{center} 
	\includegraphics[width=.57\columnwidth]{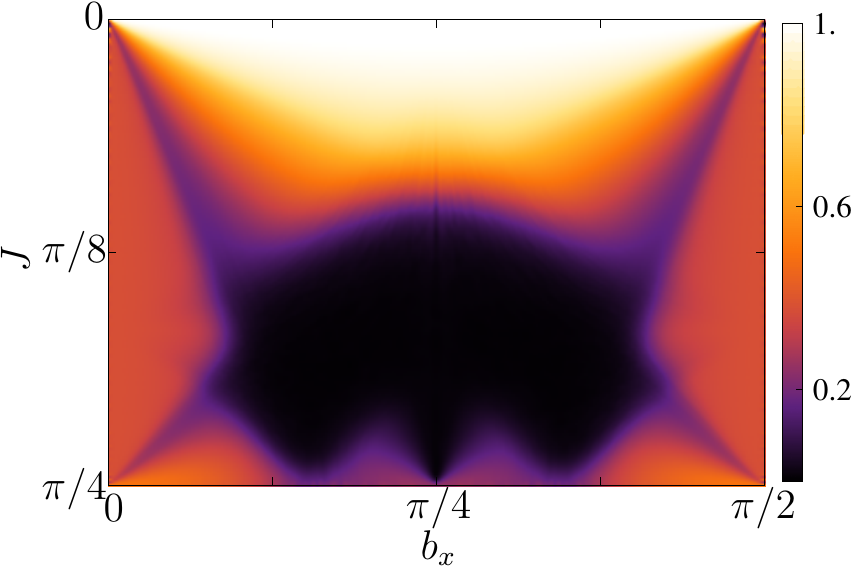}\hfill%
	\includegraphics[width=0.4\columnwidth]{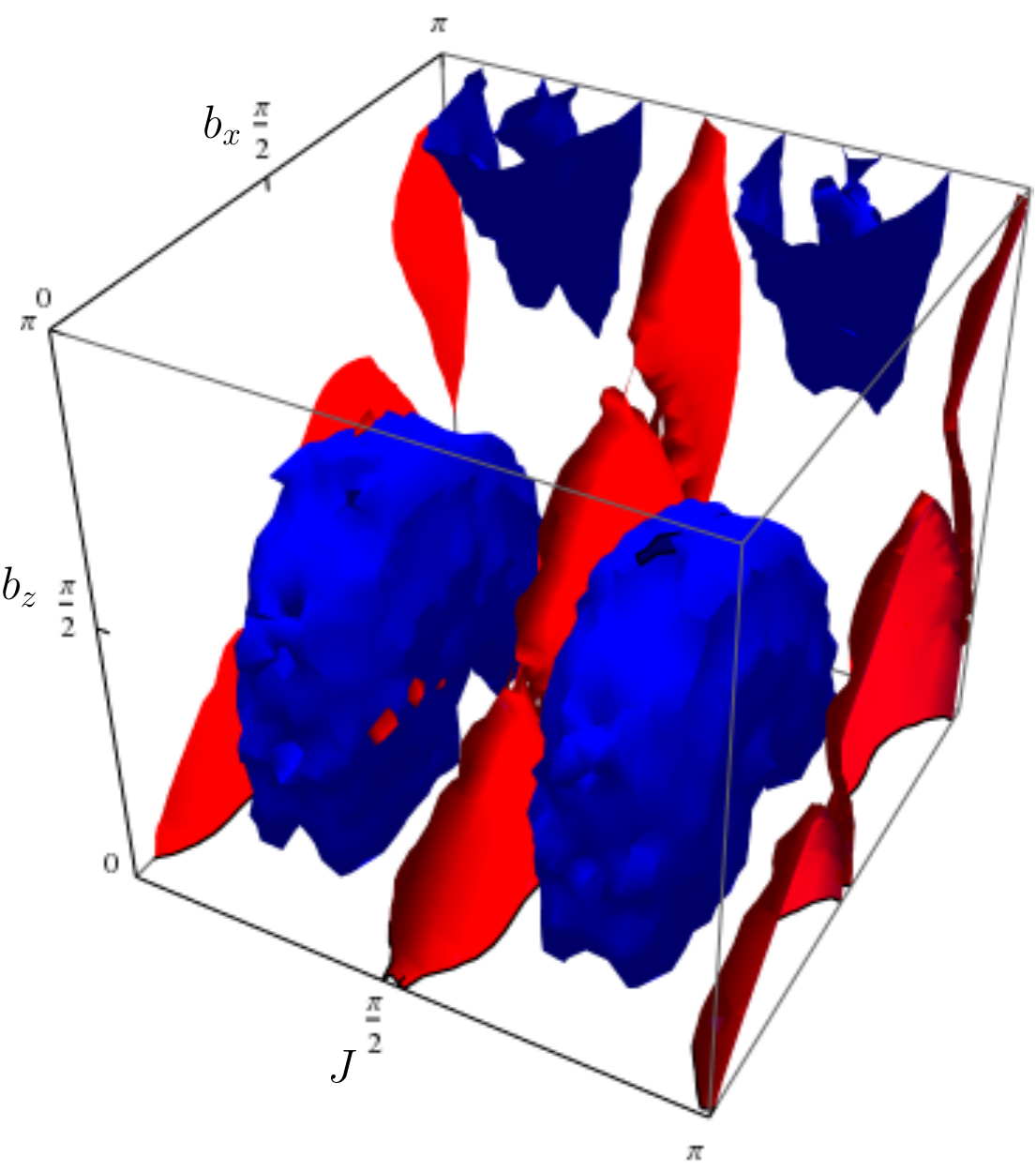}
	\end{center} \caption{(left) Correlation $\ave{C(t)}_{80}^{100}$ for
	the Ising model, for $M=S^x$, as a function of $b_x$ and $J$, with
	$M=S_x$ and $b_z=0$.   (right) Regions in which the correlations assume
	its largest and smallest values, for all the parameters of the model.
	The region within the blue surface has small correlations,
	$\ave{C(t)}_{80}^{100} \le 0.004$), whereas the region enclosed by the
	white surface has big correlation $\ave{C(t)}_{80}^{100} > 0.5$. Here,
	the size is a $4 \times 4$ lattice.
	}
	\label{fig:correlation}
\end{figure} 

In left panel of \fref{fig:correlation} we observe the dynamical susceptibility
as a function of both the transverse field $b_x$ and $J$ (here we set $b_z=0$).
There is clearly a set of parameters for which the model is not ergodic, that
is, where $\chi_M \neq 0$. However, for $J=0.5$ there seems to be a range of
$b_x$ where the correlations clearly vanish, namely for $ 0.4 < b_x < 1.2$. We
have observed also that  as we were able to increase the number of particles,
the transition was increasingly sharper.  In the right panel of
\fref{fig:correlation} we sketch the full three-dimensional phase diagram (of
order parameter $\chi_M$)  in the parameter space $(J,b_x,b_z)$, clearly
indicating distinct regions of ergodic and non-ergodic dynamics.

Comparing these data to phase diagrams of level density
\fref{fig:spectral:density}, or nearest neighbour agreement with RMT
\fref{fig:ps} one finds that there is no point to point correspondence between different regimes in the parameter space. 
One can have a flat or non-flat level density in either ergodic or non-ergodic regime for the dynamics of local observables.
This speculative conclusions is certainly surprising and calls for a deeper understanding of the role of locality of observables in long time dynamics.

\begin{figure} 
	\begin{center} 
	\includegraphics[width=.47\columnwidth]{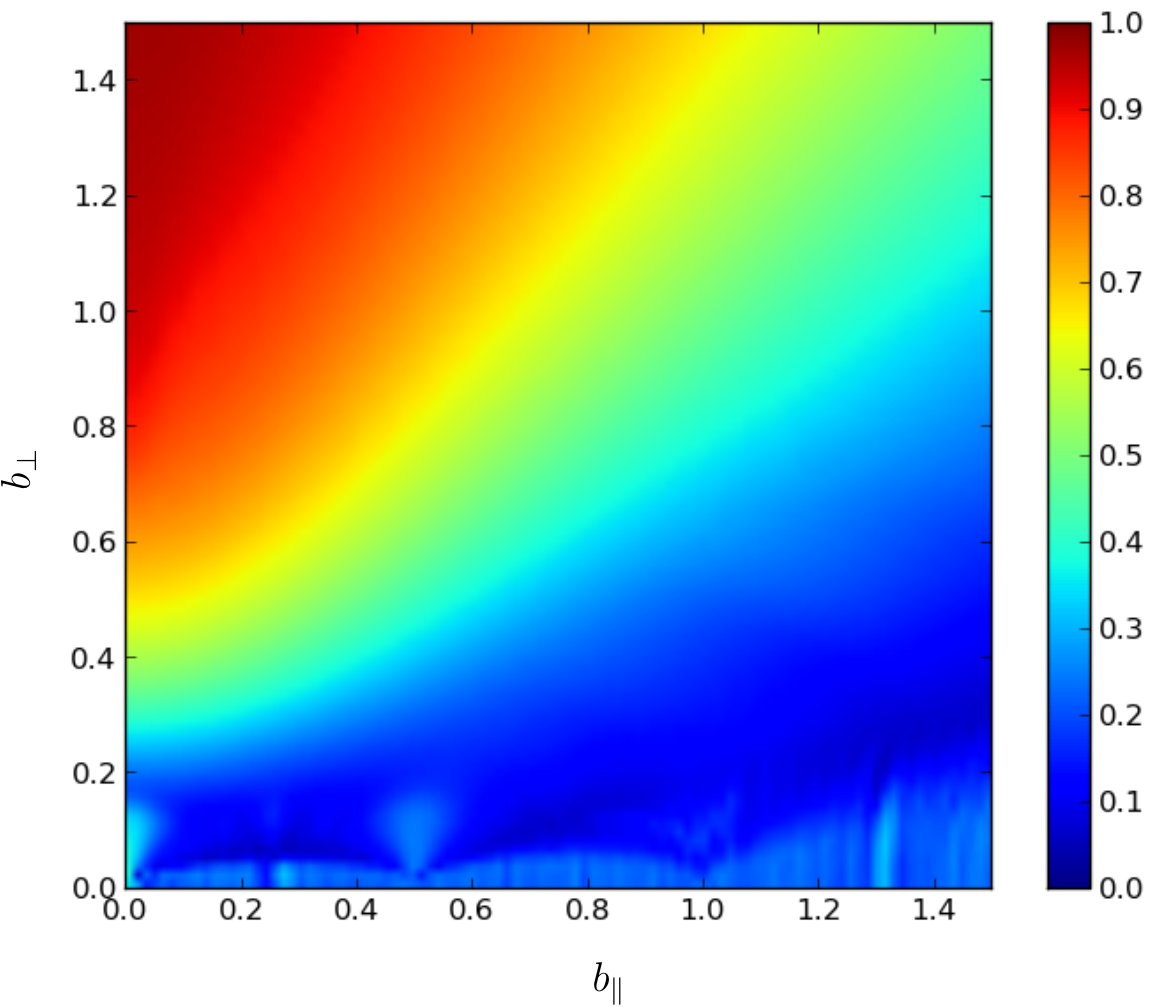} 
	\end{center}
	\caption{Correlation for the Ising model in the steady field limit, as a
	function of the components of the  field $\vec b$, fixing $J=0.25$ and
	a $5\times4$ grid.}
	\label{fig:correlacion:steady}
\end{figure} 

We have also used the same program to have a glimpse into the behavior of the
steady field model, and found hints that this is also a rich model, in which
both situations of ergodic and non-ergodic dynamics are found for different
parameter values, see
\fref{fig:correlacion:steady}.
\section{Conclusions}
\label{sec:conclusions}
In this paper we describe a computational excursion into ergodic properties of
two-dimensional periodically driven quantum spin systems. In the absence of
efficient computational techniques we implemented brute force simulation of the
system's dynamics. 
%
%
Speculating on the thermodynamic properties of the system by
inspecting an increasingly large sequence of periodic
lattices, our results suggest several rather intriguing conclusions. 
The spectral
density of the Floquet operator displays phase transitions from regions of flat
density to regions with nontrivial spectral densities dominated by nonzero
Fourier components.  Local observables display ergodic regimes with decaying
correlations and non-ergodic regimes with non-decaying correlations, which
however, do not correspond to regions of flat versus non-flat level densities.
Moreover, the level spacing distribution is essentially given by Wigner surmise
of random matrix theory over the entire parameter space, where the model is
non-integrable, and therefore, surprisingly, does not provide any useful information on system's ergodicity.  
We believe that our numerical results generate a strong
motivation for further theoretical investigations into dynamics of periodically
driven (or discrete-time) interacting spin models on 2D lattices.
 
TP acknowledges financial support by the grant P1-0044 and J1-5439 of the
Slovenian Research Agency.  Support by the  projects CONACyT 153190 and
UNAM-PAPIIT IA101713 is acknowledged by CP and EV.

\section*{References}
\bibliographystyle{unsrt}
\bibliography{article}
\begin{appendix} 
\section{Numerical implementation of the model} 

The GPU implementation is made by using Nvidia CUDA architecture for GPU
parallel computing.  We store the coefficients (in the computational basis) of
the state one wishes to evolve on the global memory of the GPU and then apply
the required quantum gates \eref{eq:floquet} on it.  The parallelization is
done, realizing that the application of a $n$-qubit gate can be decomposed in
$2^{L-n}$ independent parallel operations. Each of them shall be done by a
single thread in the GPU.

%
%

Using the threads index \verb|threadIdx.x+blockIdx.x*blockDim.x| and the
computational base we specify the entries on the state a specific thread will
compute on, for example when applying a 1-qubit gate on the second qubit, each
thread shall act on the coefficients of the components. In particular, 

\begin{align*}
\text{thread 0:} &\quad |\cdots0\underline{0}00\> , |\cdots0\underline{1}00\> \\
\text{thread 1:} &\quad |\cdots0\underline{0}01\> , |\cdots0\underline{1}01\> \\
\text{thread 2:} &\quad |\cdots0\underline{0}10\> , |\cdots0\underline{1}10\> \\
& \vdots \\
\text{thread $2^{L-1}$:} &\quad |\cdots1\underline{0}11\> , |\cdots1\underline{1}11\> .
\end{align*}
The qubit over which the gate is acting is underlined, and is the one that ``couples''
the computational states. 
By using this scheme all gates con be computed in parallel regardless of the number of qubits
it works on. The program that
realizes these operations is publicly available in~\cite{githubcuda}.

\begin{figure} 
	\begin{center} 
	\includegraphics[width=\columnwidth]{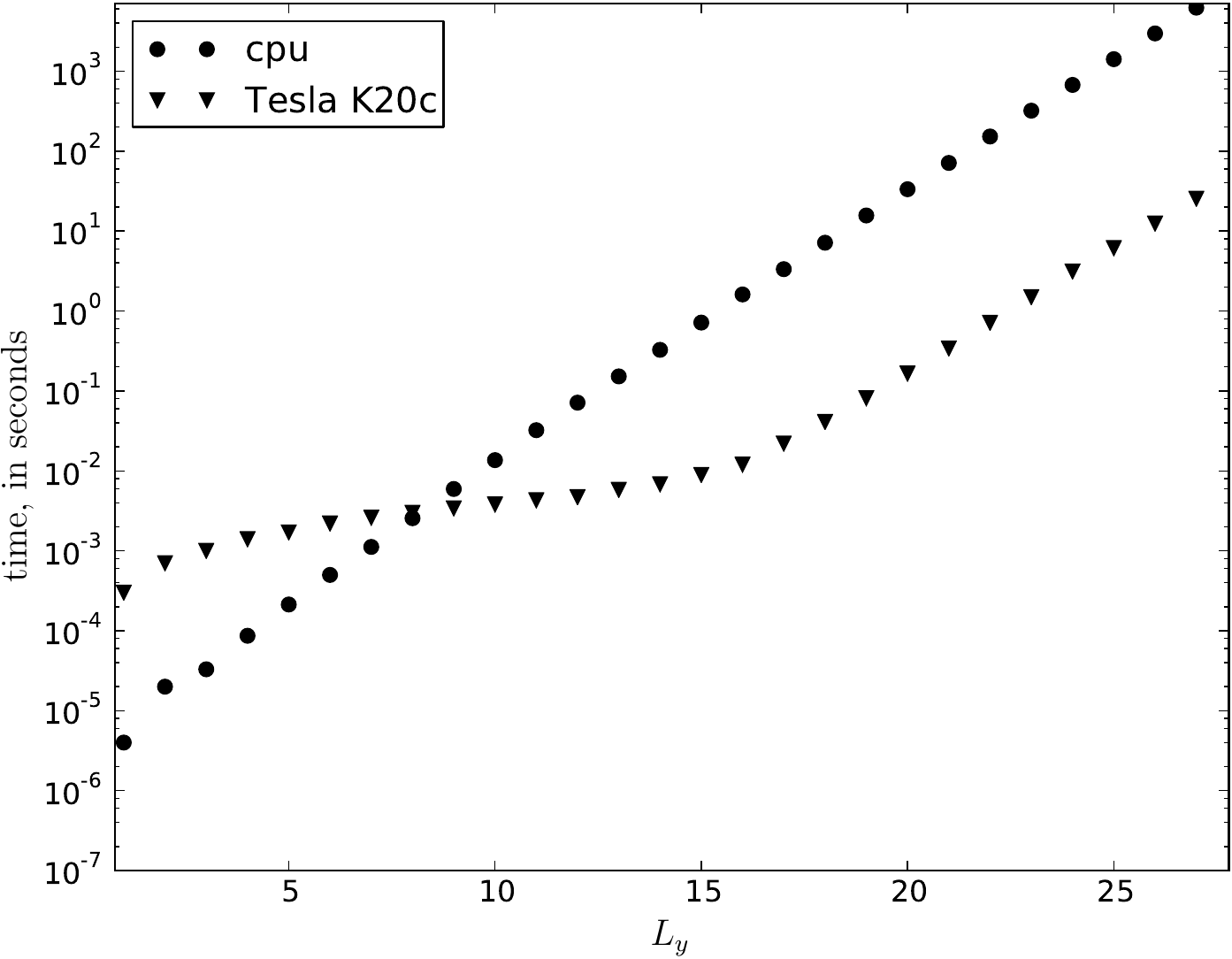} 
	\end{center}
	\caption{Time to execute a time step of the kicked Ising spin model (with
	$L_x=1$),  on a random state.  
	We compare GPU times [on a Tesla K20c
	with 5Gb of memory and 2496 Cuda cores] against a single good processor
	[AMD Opteron(TM) Processor 6212, 2600MHz]. We can see that around 13
	qubits already we have an important speed factor increase, which
	stabilizes around 240x. We are limited by the size of the memory of the
	card to 25 qubits.  } \label{fig:gpuvscpu}
\end{figure}

Let us compare the speed of the two setups, namely GPU and a usual CPU
implementation. For that we evaluate the average speed for the application of
one time step of the Kicked Ising model.  That is, the application of the
unitary operation \eref{eq:floquet} to a random state.  We apply the operator
several times for smaller system sizes, so as to get good statistics. The
results are presented in \fref{fig:gpuvscpu}, and show a more than satisfactory
speed increase.

\end{appendix}
\end{document}